\documentclass[prl, twocolumn]{revtex4}
\usepackage{amsmath}
\usepackage{amssymb}
\usepackage{graphicx}
\usepackage{algpseudocode, algorithm}
\usepackage{color}

\begin{document}


\title{Burstiness and spreading on temporal networks}


\author{Renaud Lambiotte}
\affiliation{naXys, University of Namur, Belgium}
\author{Lionel Tabourier}
\affiliation{naXys, University of Namur, Belgium}
\author{Jean-Charles Delvenne}
\affiliation{ICTEAM and CORE, Universit\'e catholique de Louvain, Belgium}

\begin{abstract}
We discuss how spreading processes on temporal networks are impacted by the shape of their inter-event time distributions. Through simple mathematical arguments and toy examples, we find that the key factor is the ordering in which events take place, a property that tends to be affected by the bulk of the distributions and not only by their tail, as usually considered in the literature. We show that a detailed modeling of the temporal patterns observed in complex networks can change dramatically the properties of a spreading process, such as the ergodicity of a random walk process or the persistence of an epidemic.
\end{abstract}


\maketitle


\section{Introduction}

When modeling diffusive processes in systems made of interacting elements \cite{Boccaletti2006}, a majority of works has adopted a Poisson viewpoint, where stochastic events take place at a constant rate. For instance, in models of disease spreading over contact networks \cite{Hethcote2000}, it is usually assumed that the probabilities per unit time of disease transmission and of  recovery from disease are constant, implying exponential distributions of the time intervals between events. This assumption has important consequences on the mathematical nature of the models, as they become memoryless and they conveniently reduce to ordinary differential equation models. In situations when only the average rate of the events is known, this approach is statistically justified by the principle of maximum entropy \cite{Jaynes1957}. However, growing evidence shows that inter-event time statistics significantly deviates from Poisson processes in a variety of systems \cite{Barabasi2012}, with important consequences on the dynamical and asymptotic properties of spreading models \cite{Rocha2010,Karsai2011,Peruani2011,Starnini2012}. For instance, the distribution of duration of most diseases has a sharp peak about the average value and is highly non-exponential \cite{Sartwell1950,Blythe1988}. Similarly, times between contacts \cite{Isella2011} or communication \cite{Kleinberg2003,Eckmann2004} between individuals also tend to deviate from a Poisson process, but this time by exhibiting a bursty behaviour, namely an intermittent switching between periods of low activity and high activity, and a fat-tailed inter-event time distributions \cite{Barabasi2005}. 

Incorporating non-exponential distributions into a mathematical modeling of diffusion leads to integro-differential equations
\cite{Hethcote1980,Keeling1997,Hoffmann2012}, where the evolution of the system at some time depends on an integration of its states over its past, i.e. the random process becomes non-Markovian. An important aspect of these equations is that they account for the importance of the time ordering of events on dynamics. Typically, for one event to take place, some other event should not have taken place before. In the case of disease spreading, for instance, an infected individual can only transmit the disease at a certain time if is has not recovered at that time. The main purpose of this article is to investigate the properties of inter-event time distributions that affect spreading. In particular, we will emphasize the importance of time ordering for two different types of diffusive processes, namely random walks and epidemic spreading.

\section{Diffusion and time ordering}

\subsection{From data to models}

In a majority of empirical systems, networks are not static entities, as edges and nodes can appear and disappear in time. A natural framework to study time-dependent complex systems is to use temporal networks \cite{Holme2012}, in which one accounts for the timings of interactions instead of assuming static connectivity. The modeling of temporal networks requires one finding the right level of abstraction, which allows for a mathematical analysis while preserving key properties of the data. In that direction, a promising approach consists in building stochastically evolving networks where the appearance of edges between nodes is a stochastic process, built such as to preserve the inter-event time distribution observed in the data. 

In practice, stochastic models are constructed as follows. Let us consider an empirical system observed during a time interval $T$, made of $N$ nodes, and where edges between two nodes, say from $i$ to $j$, appear at times $t_{ij}=\{t_{ij}^{(1)},t_{ij}^{(2)},...,t_{ij}^{(n_{ij})} \}$, where $n_{ij}$ is the total number of activations of that edge. The sequence is ordered such that $t_{ij}^{(a)}<t_{ij}^{(b)}$ if $a<b$. For the sake of simplicity, we will further assume that edges remain present for infinitesimally small times, in order to avoid several edges to be present at the same time. The modeling step consists in replacing the exact sequence of activation times by a random sequence where events take place according to an inter-activation time $f_{ij}(\tau)$ fitted on the data. More precisely, $f_{ij}(\tau) d\tau$ is the probability to observe a time interval of duration in $[\tau,\tau+d\tau]$  between two activations of the edge. By convention, $f_{ij}(\tau)$ is only defined for edges that are activated at least once. For those, $f_{ij}(\tau)$ verifies 
\begin{equation}
\int_0^\infty f_{ij}(\tau) d\tau = 1,
\end{equation}
and 
\begin{equation}
\int_0^\infty \tau f_{ij}(\tau) d\tau = \langle \tau \rangle_{ij}
\end{equation}
gives the expected time between two activations of an edge.

When modeling the diffusion of an entity on the network, however, the distribution $f_{ij}(\tau)$ only plays an indirect role. The  important quantity is instead the waiting time distribution $\psi_{ij}(t)$ that the entity arriving on $i$ has to wait for a duration $t$ before an edge towards $j$ is available. The waiting time $t$ is often called relay time. In epidemic spreading, it is the time it takes for a newly infected node to spread the infection further via the corresponding link.
Assuming that the activations of neighbouring edges are independent \footnote{For a discussion of the case of correlated events, we refer to \cite{Kivela2012}} and  that nodes become infected at uniformly random times, inter-activation time distribution and  waiting time distribution verify the relation \cite{Miritello2011,Kivela2012}
\begin{equation}
\psi_{ij}(t) = \frac{1}{\langle \tau \rangle_{ij}}\int_t^{\infty}  f_{ij}(\tau) d\tau
\end{equation}
derived as follows. The probability for the walker to arrive in an inter-activation time of duration $\tau$ is proportional to $\tau$, more precisely
$\frac{\tau f_{ij}(\tau)}{\langle \tau \rangle_{ij}}$. 
The probability to wait for a duration of length $t$ is simply given by the probability to land in an inter-activation time of length $\tau \geq t$, precisely at time $\tau - t$ in this interval. As the probability to arrive in an interval is uniform, one finds
\begin{equation}
\psi_{ij}(t) = \int_t^\infty  \frac{\tau f_{ij}(\tau)}{\langle \tau \rangle_{ij}} \frac{1}{\tau} d\tau 
\end{equation}
and hence the above relation. 
The average waiting time can be computed from the latter after integrating by parts
\begin{equation}
\langle t \rangle_{ij} = \int_0^\infty t\psi_{ij}(t) dt = \frac{1}{2} \frac{\langle \tau^2\rangle_{ij}}{\langle \tau\rangle_{ij}}
\end{equation}
showing that the average waiting time depends on the variance of the inter-activation time. At a fixed value of the average inter-activation time, the waiting time can be arbitrarily large if the variance of inter-activation times is sufficiently large.
This paradox, often called waiting time paradox or bus paradox in queuing theory \cite{Allen1990}, is an example of length-biased sampling. Let us note that waiting-times and inter-activation times have the same distribution when the process is Poissonian, in which case
\begin{eqnarray}
\psi_{ij}(t) = f_{ij}(t) =\frac{1}{\langle t\rangle_{ij}} \exp \left( -\frac{1}{\langle t\rangle}_{ij} \right) 
\end{eqnarray}
and that their tail has the same nature in the case of power-law tails
\begin{eqnarray}
\psi_{ij}(t) \sim t^{-\alpha} &\Leftrightarrow& f_{ij}(\tau) \sim \tau^{-(\alpha+1)}.
\end{eqnarray}
The quantity $\langle t \rangle_{ij}/\langle \tau\rangle_{ij} -1$, which is zero for a Poisson process and positive or even infinite for a power law, is a standard measure for the burstiness of a process \cite{Goh2008, Kivela2012}.

\subsection{Two spreading models}

In this section, we focus on two popular models for diffusion on networks, and show that in both cases, the non-Markovianity of the random process alters the diffusion by changing the ordering in which two types of events take place. The nature of these events is however different in each case. From now on, we will only consider the waiting-time distribution $\psi_{ij}(t)$, relevant to describe dynamical processes on networks, and focus on the associated stochastic temporal network,  in which edges appear randomly according to the assigned waiting times.

\subsubsection{Random Walks}

Random walk processes are a generic model for diffusion, also used to uncover prominent structural features of networks. Applied to stochastic temporal networks, the model is defined in continuous time as follows \cite{Hoffmann2012}. A walker located at a node $i$ remains on it until an edge leaving $i$ toward some node $j$ appears. When such an event occurs, the walker jumps to $j$ without delay and then waits until an edge leaving $j$ appears. It is important to note that the probability for the walker to jump to $j$ depends on $\psi_{ij}(t)$, but also on all $\psi_{ik}(t)$, where $k$ are neighbours of $i$, 
because the walker takes the first edge available for transport. Once a walker has left a node, edges leaving this node become useless for transport. For this reason, the probability to actually make a step from $i$ to $j$ is given by
\begin{align} 
 \label{eq:effective0}
	T_{ij}\left(t\right)  =\psi_{ij}\left(t\right)\times\prod_{k\neq j} \int_{t}^{\infty}\psi_{ik}\left(t'\right)dt'
\end{align} 
where  each factor in the product denotes the probability that an edge does not appear before time $t$.
The probability for making a jump to node $j$ is given by the effective transition matrix
\begin{equation} \label{eq:effective}
	\mathbb{T}_{ij} \equiv \int_0^\infty T_{ij}(t) dt\,,
\end{equation} 
that verifies 
\begin{equation} \label{eq:effective1}
\sum_j	\mathbb{T}_{ij} =1.
\end{equation} 
When only two edges leave node $i$, say to $j$ and $k$, (\ref{eq:effective0}) simplifies into
\begin{equation}
\label{eq:rw1}
	T_{ij}\left(t\right)=\psi_{ij}\left(t\right) \int_{t}^{\infty}\psi_{ik}\left(t'\right)dt'. 
\end{equation}

\subsubsection{Epidemic spreading}

Epidemic spreading differs from random walk processes because the number of infected individuals is not conserved.  It may decrease when an infected person recovers, or increase when an infected person infects several of its contacts. When applied on stochastic temporal networks, standard models of epidemic spreading are characterized by two distributions: i) the probability distribution $\psi_{ij}(t)$ that the infected node $i$ makes a contact sufficient to transmit the
disease to node $j$ at time $t$, after he has been infected at time $0$; the probability distribution $r_i(t)$ that node $i$ infected by the disease recovers at time $t$.
As an infected individual can only transmit the disease to a susceptible
neighbor if it is still infected at the time of contact \cite{Karrer2010}, the probability of transmission from $i$ to $j$, at time $t$ after $i$ has been infected is given by
\begin{equation}
P_{ij}(t) = \psi_{ij}(t) \int_t^\infty r_i(t') dt'.
\label{eq:ep1}
\end{equation}
The overall probability that node $i$ infects node $j$ before it recovers is given by
\begin{equation}
\mathbb{P}_{ij} = \int_{0}^{\infty}  P_{ij}(t) dt.
\label{eq:ep2}
\end{equation}
This quantity is usually referred to as the \textit{transmissibility} or \textit{infectivity} of the disease.

\subsection{Effect of the shape of the distribution}

A comparison between  (\ref{eq:rw1}) and (\ref{eq:ep1}) clearly shows the similarities and differences between the corresponding dynamical processes. In both cases, it is the overall probability that edge $ij$ appears before another type of event that determines  
if the spreading goes through this edge or not.
In the case of a random walk, it is a comparison with the activations of neighbouring edges that matters. In the case of epidemic spreading, it is instead a comparison with the recovery process. $\mathbb{T}_{ij}$ and $\mathbb{P}_{ij}$ determine the importance of an edge for each random process and thus the pathways of diffusion. These quantities provide a static representation of the dynamical process, not sufficient to determine properties such as the speed of propagation, but allowing to predict some of their their asymptotic properties. For random walks, $\mathbb{T}_{ij}$ defines a standard Markov chain that determines the asymptotic properties of the continuous-time random walk. For epidemic spreading, $\mathbb{P}_{ij}$ directly affects the basic reproduction number $R_0$, namely the average number of additional people that a person infects before recovering, in the limit when a vast majority of the population is susceptible. The point $R_0=1$ defines the epidemic threshold separating between growing and decreasing spreading. In tree-like networks, where all nodes have the same transmissibility $\mathbb{P}$, one finds $R=\mathbb{P} \langle k (k-1)\rangle/\langle k \rangle$, where $ \langle k (k-1)\rangle/\langle k \rangle$  is the 
expected number of susceptible neighbors of an infected node. The epidemic threshold is thus reduced either by reducing the transmissibility or the ratio $ \langle k^2\rangle/\langle k \rangle$.

In epidemiology, researchers have mainly focused on the non-exponential nature of the recovery time distribution, as infectious periods tend to be closely centered on the mean duration of infection. $r_i(t)$ is usually modeled by gamma distributions \cite{Wearing2005} or approximated by delta peaks. Let us also note that when modeling epidemic spreading of ideas or trends in social networks, fat-tailed distributions are more appropriate \cite{Sornette2004}. In complex systems theory and computer science, research has focused on the waiting time distribution $\psi_{ij}(t)$, either modeled by power-law \cite{Barabasi2005}, Weibull \cite{Mieghem2013} or log-normal distributions \cite{Malmgren2008} to account for the bursty dynamics observed in the empirical data. In this direction, it is interesting to point that much research has studied the effect of the tail, typically a power-law tail, of the distribution on spreading. Yet, it is not the shape of the tail, nor the moments of the distribution, that affect the pathways of diffusion.  What matters is instead the relative position of one distribution with another distribution, as Eqs. (\ref{eq:rw1}) and (\ref{eq:ep1}) clearly show. For an edge to be important, it should appear often before some other random event. It is true that a small average or a fat tail \footnote{When considering two distributions with the same average, a distribution with a fatter tail tends to have more probability assigned to small values of the random variable.} tend to favor a distribution, but it is not always the case as the following toy example clearly shows. 

Let us consider epidemic spreading on a regular tree of identical nodes with degree $3$. Each node has the recovery distribution $r(t)=\delta(t-1)$, e.g. recovery times occur exactly at the average value $1$, and
each edge is characterized by the waiting time distribution 
\begin{eqnarray} \label{eq:ex}
 \psi(t) = \begin{cases}
\alpha    & {\rm for} \quad t<1,\cr
\frac{1-\alpha}{t^2}      & {\rm for} \quad t\geq1,
\end{cases} 
\end{eqnarray}
where $\alpha \in [0,1]$ tunes the shape of the distribution.
For any value of $\alpha$: i) the distribution is properly normalized; ii) its average (hence burstiness) is infinite; iii) it exhibits a power-law tail with exponent $2$. Despite sharing these properties, the transmissibility of an edge continuously varies between $0$ and $1$ when varying $\alpha$, as
\begin{equation}
\mathbb{P}=  \int_{0}^{\infty} \psi(t) \int_{t}^{\infty} \delta (t'-1) dt' dt = \int_{0}^{1}  \psi(t) dt = \alpha.
\end{equation}
This observations implies qualitatively and quantitatively different spreading behaviours when tuning $\alpha$,
as the system is above the epidemic threshold when $\alpha> 1/2$, and below otherwise. 

A corresponding example can be found for random walks. Consider a node from which leave two edges whose waiting times distributions follow (\ref{eq:ex}) with parameters $\alpha_1$ and $\alpha_2$. The resulting transition probabilities can be spread along those edges in any possible way by a suitable choice of $\alpha_1$ and $\alpha_2$, leading to possibly dramatically different behaviours of the random walker. For example if one of the edges is the only bridge between two parts
of the graph, modulating its transition probability to zero can ultimately disconnect the graph entirely, preventing the ergodicity of the random walker.

Before closing this section, let us mention a couple of interesting properties of (\ref{eq:rw1}) and (\ref{eq:ep1}). In general, these equations define the overall probability that an event $A$ takes place before some other event $B$
\begin{equation}
\label{xx}
p_A= \int_{0}^{\infty} a(t) \int_t^\infty b(t')  dt' dt,
\end{equation}
where $a(t)$ and $b(t)$ are two probability distributions. The overall probability that $B$ takes place before $A$ is similarly defined
 \begin{equation}
p_B= \int_{0}^{\infty} b(t) \int_t^\infty a(t')  dt' dt,
\end{equation}
and it is straightforward to show that $p_A+p_B=1$.
When both distributions are identical, $a(t)=b(t)$, one finds $p_A=p_B=1/2$ as expected due to symmetry reasons.
When both distributions are exponentials,  $a(t)=r_A e^{-r_A t}$ and $b(t)=r_B e^{-r_B t}$, where $r_A$ and $r_B$ are the rates at which events take place, one finds 
\begin{equation}
\label{xxx}
p_A=\frac{r_A}{r_A+r_B}.
\end{equation}
 Finally, when one type of event is exponentially distributed, say $b(t)=r_B e^{-r_B t}$, (\ref{xx}) simply amounts to the Laplace transform in the variable $r_B$ of the other distribution
\begin{equation}
p_A= \int_{0}^{\infty} a(t) e^{-r_B t} dt.
\end{equation}

\section{Discussion}

The main purpose of this paper was to identify the properties of temporal patterns of edges and nodes that affect pathways of diffusion on time-evolving networks. This problem has attracted much attention in recent years, often leading to claims that temporal heterogeneity can significantly alter spreading. However, there is still no general understanding of the mechanisms by which burstiness affects the diffusive process. Is it the tail of the inter-event time distribution that matters, as often suggested, or its variance? Our work suggests that the most important factor is instead the time-ordering of events, which identifies the importance of an edge as the overall probability (\ref{xx}) that it appears before some other event takes place;  the nature of those competing events depends heavily on the kind of spreading process under scrutiny. 
This measure of dynamical weight depends on the full probability distributions of those competing events, and seemingly more critically on their bulks than on their tails, because the probability mass is mainly concentrated in the bulk. Importantly, (\ref{xx}) is a scalar measure of importance of edges that  aggregates their full sequence of activation, and thus provides a static picture properly taking into account the temporal dynamics of edges. Contrary to standard procedures, the importance of an edge is in general not proportional to its number of activations, as in (\ref{xxx}), but to the probability that it participates in the diffusive process. Future work will focus on the transient properties of the diffusive processes, and aim at evaluating the effect of inter-event time distributions on properties such as the mixing time, or the peak time.

\section*{Acknowledgment}
RL acknowledge financial support from F.R.S-FNRS and from the COST Action TD1210 KnowEscape. JCD acknowledges support from the Action de Recherche Concerte `Large graphs and networks' funded by the Federation Wallonia-Brussels. This paper presents research results of the Belgian Network DYSCO (Dynamical Systems, Control, and Optimization), funded by the Interuniversity Attraction Poles Programme, initiated by the Belgian State, Science Policy Office.

\end{document}